# Systematic review of self-supervised foundation models for brain network representation using electroencephalography

Hannah Portmann and Yosuke Morishima

Translational Research Center, University Hospital of Psychiatry and Psychotherapy, University of Bern

Corresponding author

Yosuke Morishima

Translational research center

University Hospital of Psychiatry and Psychotherapy

University of Bern

Bolligenstrasse 111, 3000 Bern, Switzerland

yosuke.morishima@unibe.ch

# Abstract

Automated analysis of electroencephalography (EEG) has recently undergone a paradigm shift. The introduction of transformer architectures and self-supervised pretraining (SSL) has led to the development of EEG foundation models. These models are pretrained on large amounts of unlabeled data and can be adapted to a range of downstream tasks. This systematic review summarizes recent SSL-trained EEG foundation models that learn whole-brain representations from multichannel EEG rather than representations derived from a single channel. We searched PubMed, IEEE Xplore, Scopus, and arXiv through July 21, 2025. Nineteen preprints and peer-reviewed articles met inclusion criteria. We extracted information regarding pretraining datasets, model architectures, pretraining SSL objectives, and downstream task applications. While pretraining data heavily relied on the Temple University EEG corpus, there was significant heterogeneity in model architecture and training objectives across studies. Transformer architectures were identified as the predominant pretraining architecture with state-space models such as MAMBA and S4 as emerging alternatives. Concerning SSL objectives, masked auto-encoding was most common, and other studies incorporate contrastive learning. Downstream tasks varied widely and implemented diverse fine-tuning strategies, which made direct comparison challenging. Furthermore, most studies used single-task fine-tuning, and a generalizable EEG foundation model remains lacking. In conclusion, the field is advancing rapidly but still limited by limited dataset diversity and the absence of standardized benchmarks. Progress will likely depend on larger and more diverse pretraining datasets, standardized evaluation protocols, and multi-task validation. The development will advance EEG foundation models towards robust and general-purpose relevant to both basic and clinical applications.

# Introduction

Electroencephalography (EEG) has provided a non-invasive window into human brain function, since Hans Berger's initial recording in 1924 (Berger, 1929). Over the last century, EEG has become a core tool in clinical neurology and psychiatry as well as basic neuroscience (Salisbury et al., 2024; Brigo and Mecarelli, 2025). It has revealed oscillatory neural dynamics, event-related responses, and resting-state fluctuations, and it has provided diagnostic tools for epilepsy and sleep disorders (Brigo and Mecarelli, 2025). Beyond diagnostics, EEG is increasingly utilized to "decode" internal cognitive states for brain-computer interface (BCI) applications (Abiri et al., 2019; Rashid et al., 2020; Erat et al., 2024; Jaipriya and Sriharipriya, 2024). However, EEG signals are inherently non-stationary, susceptible to noise, and exhibit high inter-subject variability, making manual interpretation and automated analysis significant challenges (Wei et al., 2018; Saha and Baumert, 2020).

Historically, automated EEG analysis relied on classical machine learning classifiers, such as support vector machines (SVM) and random forests (Lotte et al., 2018; Aggarwal and Chugh, 2022). These approaches typically require extensive feature engineering—extracting handcrafted features like power spectral density (PSD), differential entropy (DE), or common spatial patterns (CSP). While effective for specific tasks, these methods are limited by their reliance on domain expertise and their low capability to generalize well across different subjects or tasks without manual recalibration (Wei et al., 2018; Saha and Baumert, 2020).

Deep learning introduced a paradigm shift, allowing models to learn features directly from raw data (Craik et al., 2019; Roy et al., 2019). Convolutional neural networks (CNNs) (Krizhevsky et al., 2012), adapted from their success in computer vision, became a widely-used architecture for EEG. Models such as EEGNet (Lawhern et al., 2018) and DeepConvNet (Ding and He, 2017) successfully integrated spatio-temporal feature extraction through convolutional layers. However, while CNNs have an excellent capability to capture local patterns within short time windows (e.g., 1–4 seconds), they often struggle to capture long-range temporal dependencies critical for understanding complex brain dynamics.

To address temporal continuity, recurrent neural networks (RNNs), and specifically long short-term memory (LSTM) networks, were introduced in EEG applications (Hochreiter and Schmidhuber, 1997). LSTMs utilize a gating mechanism to maintain longitudinal information through a cell state, mitigating the vanishing gradient problem inherent in conventional RNNs. LSTM networks combined with CNNs have shown stronger performance for sequence modeling tasks such as sleep staging (Chen et al., 2020; Efe and Ozsen, 2023; Zhu et al., 2025) or seizure detection (Xu et al., 2020; Kashefi Amiri et al., 2025) when trained on large-scale datasets.

A paradigm shift occurred with the introduction of the transformer architecture (Vaswani et al., 2017), which overcame the sequential processing limitations of RNNs. Driven by the success of Large Language Models (LLMs), this shift relies on two key ideas: self-supervised learning (SSL) for pretraining and advanced sequence modeling architectures for long-range temporal dependencies. These two key advancements have given rise to "Foundation Models"—large-scale models pretrained on vast amounts of unlabeled data that can be fine-tuned for diverse downstream tasks (Bommasani et al., 2022).

First, SSL mitigates the scarcity of labeled medical data by using the data itself as the supervisory signal (Albelwi, 2022; Liu et al., 2023; Rani et al., 2023). Learning objectives of SSL are often roughly categorized into two main types (Liu et al., 2023). The first learning objective is contrastive learning, which generates synthetic positive and negative data from the original data and learns parameters to minimize the distance between the original and positive samples, while maximizing the distance to negative samples. The second is masked prediction, where a part of the sequence is masked and predicted by surrounding data. For the application of SSL for EEG data (Rafiei et al., 2024; Weng et al., 2025), predicted data can be either original EEG signals, EEG spectra, or tokenized data. This can be further subcategorized into random masking and sequential masking (i.e. autoregressive tasks). In this way, models can learn robust representations from large-scale unlabeled datasets, overriding the bottleneck of tedious manual annotation.

Second, advanced sequence modeling allows a model to capture global, time-distant relationships within the signal. Unlike CNNs or RNNs, the transformer architecture effectively captures the state-dependent and non-stationary nature of EEG (Abibullaev et al., 2023; Vafaei and Hosseini, 2025). Its core advancement, the attention mechanism, adaptively weighs the importance of different signal segments regardless of their distance in time.

Building on these advances, a new generation of EEG "foundation models" has begun to emerge. In this context, foundation models are pretrained in a self-supervised fashion on large, heterogeneous EEG corpora and subsequently adapted to a wide range of downstream tasks. Early work such as BENDR demonstrated that contrastive self-supervised pretraining on a large-scale EEG dataset enables a single model to transfer across datasets and tasks, outperforming task-specific models in several BCI applications (Kostas et al., 2021).

Despite its capabilities, the transformer architecture however suffers from quadratic computational complexity regarding sequence length, complicating the analysis of long-duration EEG recordings. To mitigate this, recent research has adopted State-Space Models (SSMs) such as MAMBA (Gu and Dao, 2024) and structured state space sequence modelling (S4) (Gu et al., 2022). MAMBA has introduced the selection mechanism, which adaptively changes parameters for every token, while S4 introduced convolution to compute long context. These architectures offer global context modeling with linear complexity, representing the premise of efficient and long-context EEG foundation models.

Given this rapid technological evolution, the present systematic review aims to provide a synthesis by focusing specifically on EEG foundation models trained with self-supervised learning objectives. We survey the state of the field in terms of (i) model architectures and SSL objectives; (ii) scale, diversity, and curation of pretraining datasets; (iii) downstream tasks and evaluation metrics; and (iv) reported benefits in terms of label efficiency, cross-subject and cross-dataset generalization, and robustness. For completeness and to capture the most recent developments in this rapidly evolving area, we include both peer-reviewed articles and preprints. Taken together, this review seeks to clarify the landscape of EEG foundation models based on self-supervised learning and to identify key methodological and translational challenges that future works must address.

## Methods

**Search strategy and selection criteria**

To identify relevant articles, we searched Arxiv, Scopus, Medline, and IEEExplore on July 21, 2025. We set the oldest cut-off the search in 2017, as the first transformer for NLP study (Vaswani et al., 2017) was published in 2017. We used the following set of terminology for comprehensive coverage of relevant articles after preliminary search. The sets of keywords were the following. First, defining EEG data modality. The second defines model architecture, the third defines learning objectives and foundation models. We had to use long and less structured sets of key words due to unstructured key words settings in Arxiv search.

("electroencephalography" OR "EEG" OR "electroencephalogram") AND

("transformer" OR "self-attention" OR "attention mechanism" OR "attention-based" OR "encoder-decoder" OR "BERT-style" OR "GPT-style" OR "vision transformer" OR "ViT" OR "graph attention" OR "Mamba")

( "foundation model" OR "large-scale model" OR "pre-train" OR "pretrain" OR "self-supervised" OR "self supervision" OR "unsupervised representation learning" OR "representation learning" OR "masked autoencoder" OR "MAE" OR "masked signal model" OR "MSM" OR "masked reconstruction" OR "masked recovery", "reconstruction-based" OR "contrastive learning" OR "contrastive pre-train" OR "contrastive predictive coding" OR "CPC" OR "relative positioning" OR "temporal shuffling" OR "instance discrimination" OR "SimCLR" OR "MoCo" OR "momentum contrast" OR "Barlow Twins" OR OR "BYOL" OR "multimodal" OR "cross-modal" OR "language-signal" OR "EEG-to-text" OR "graph neural network" OR "GNN")

We included studies based on the following inclusion criteria.

- Studies were published in English.
- A self-supervised learning task was used for pretraining
- Models were pretrained on large scale datasets
- Studies examined cross-task generalization

Exclusion criteria were specifically set to exclude studies using only one or two EEG channels, intended to capture EEG characteristics without considering whole brain representations. Two authors (HP and YM) independently assessed all studies retrieved through title and abstract screening, full-text review and data extraction. Then, they arbitrated in cases of disagreement.

## Data extraction

Authors HP and YM independently extracted data from all included studies. Information extracted from the studies is followings: dataset size, dataset task domain, number of channels and sampling rate used for model training, whether and how spatial information was integrated, model architecture, model size, self-supervised training task, and downstream tasks.

## Results

The literature search initially identified 1,343 records. After removing duplicates, 880 unique studies remained (Figure 1). After full-text evaluation, 19 studies met the inclusion criteria (Table 1).

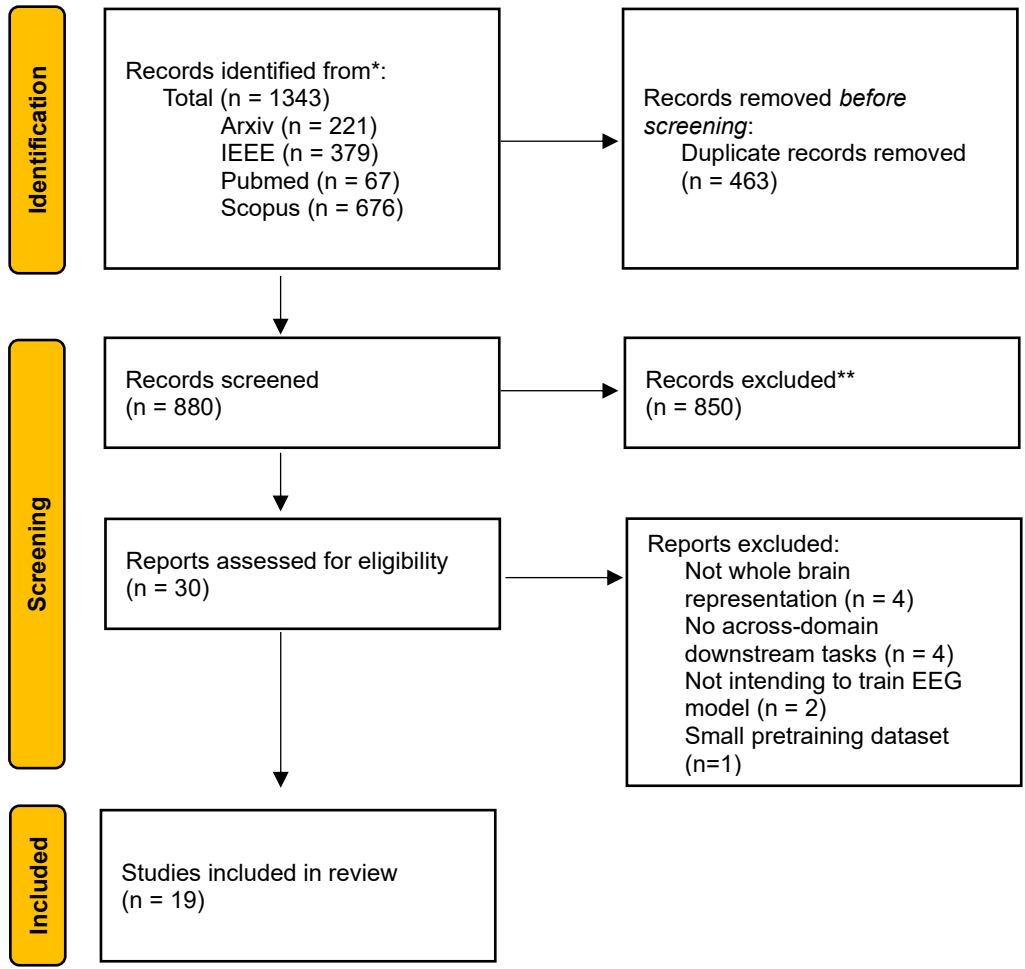

Figure 1 Prisma flowchart of the systematic review

## Dataset

Data sources varied significantly across the included studies (Table 1). Eleven studies used a single large-scale EEG dataset, mostly the Temple University Hospital EEG (TUEG) corpus (Obeid and Picone, 2016). In total, 12 studies used the TUEG corpus. The remaining studies incorporated open brain–computer interface (BCI) datasets, while a single study used a proprietary, self-acquired dataset. Four studies used between two and six datasets, while four studies used more than 10 datasets. The total size of the pretraining data was often more than 2,000 hours, and five studies used more than 10,000 hours of data for pretraining.

The number of channels used for model training ranged from eight to 128. Almost half of the studies had variable numbers of channels depending on the data used for model pretraining, while the remaining studies pretrained models on a fixed set of channels for pretraining.

Regarding the preprocessing of the pretraining data, we surveyed filtering and sampling rates, which are crucial for understanding signal characteristics and reducing computational complexity. In addition, scalp EEG contains relatively little information in the high-gamma range (>100 Hz). Regarding filtering, 12 studies applied band-pass filtering. The low cut-off frequency was often set at either 0.1 or 0.5 Hz, and the high cut-off frequency was mostly set at 100 Hz. Three studies applied low-pass filtering, and four studies did not provide filtering information. Regarding the sampling rate of the EEG data used for pretraining , EEG data were typically downsampled to reduce computational load. Sixteen out of 19 studies provided the downsampled sampling rate for pretraining, which mostly varied between 125 and 256 Hz, except for a single study that adopted 500 Hz.

**Model architecture**

Seventeen out of 19 models used a transformer-based backbone, while three models used SSMs (MAMBA and S4) (Table 2). Transformer backbone models can be categorized into 1) Vanilla transformers (4 models), 2) vision transformers (3 models), 3) CNN + transformers (6 models), or 4) vector quantization tokenizer + transformers (4 models).

Among the models utilizing a transformer backbone, attention mechanisms were implemented in different ways. Six models adopted attention in the temporal dimension, while one only in the spatial dimension. The remaining ten models implemented spatiotemporal attention: six implemented joint spatiotemporal attention, and four implemented spatial and temporal attention separately.

Neural representation of EEG signals involves both local and global features. Thus, the location of electrodes matters for understanding whole-brain representation. Fourteen out of the 19 models implemented mechanisms to encode the spatial information of EEG electrodes. Either fixed (3 models) or learnable channel embeddings (7 models) were widely used. Two studies implemented positioning with respect to a head model. Three models used a learnable 2D convolution.

Model size also varied significantly among studies. The number of trainable parameters was reported in 14 out of 19 studies. The number of trainable parameters varied from 3 million to 540 million parameters, and some studies provided several model sizes by varying MLP layers and dimensions, as well as embedding dimensions.

**Pretraining objectives**

SSL objectives also varied among models (Table 2). Fourteen models used one-stage pretraining, where one pretraining stage with one loss function was used to train the model using SSL. Five of these models used a masked reconstruction objective targeting continuous signal or value prediction. Two models used an autoregressive approach, where the future signal (continuous) or token (discrete) was predicted based only on the preceding signal or tokens, with all future parts masked. Three one-stage pretraining models used contrastive learning for pretraining. The remaining four models used a combination of SSL objectives, where the separate losses were summed together to compute the final loss function. All four used continuous masked prediction and an additional specific objective, including contrastive learning, spatio-temporal representation alignment, auto-regression, and estimation of the frequency power bands.

Five models used two-stage pretraining, which included a first pretraining stage during which the EEG signals were discretized into classes or codebooks. During the second stage, the models were trained to predict the correct classes from the first stage from masked inputs (discrete masked token prediction). These differ from the one-stage pretraining in that the first stage of training was necessary to be completed before doing the second pretraining stage.

Overall, the masked reconstruction objective was used most frequently, targeting either continuous signal or value prediction (regression) (12 models) or discrete token prediction (5 models).

**Fine-tuning and downstream task**

After model pretraining, fine-tuning was applied to perform downstream tasks in all studies (Table 3). All but the ALFEE study performed single-task fine-tuning, resulting in task-specific rather than general-purpose models. In addition, 15 studies performed full-parameter fine-tuning, while 3 fine-tuned only specific modules and 12 updated only the classification head. Eight studies compared full-parameter tuning against other approaches, and full-parameter tuning performed best in seven studies.

Downstream tasks described in the included studies were diverse. Clinical tasks included seizure detection, sleep stage classification, and the detection of pathological or abnormal EEG activity. BCI and basic cognitive tasks included motor imagery and intention, motor execution, emotion recognition, and event-related potentials. Other tasks included subject identification and artifact detection tasks, such as eye-blink detection. The number of downstream tasks included also varied between one and eight tasks. Among these tasks, motor imagery (10 studies), emotion recognition (8 studies), seizure detection (8 studies), sleep staging (7 studies), and abnormal EEG event classification tasks (10 studies) were frequently used as downstream benchmarks.

# Discussion

The current systematic review has surveyed EEG foundation models designed for whole brain representations using self-supervised learning (SSL). We identified 19 studies in the literature to date. In terms of timeline, the seminal BENDR paper was published in 2021 (Kostas et al., 2021), then followed by the BIOT paper in 2023 (Yang et al., 2023). Since 2024, the number of proposed EEG foundation models has grown rapidly. Furthermore, the field has matured sufficiently to host dedicated challenges, such as the EEG Foundation Model Challenge (https://eeg2025.github.io/). Although this review does not cover models proposed during that specific competition due to the search period, it provides a comprehensive overview of how architecture was developed during the field's formative stages.

A major finding regarding pretraining data is the predominance of the (TUEG) corpus. While it offers an indispensable volume of clinical data from a single source, relying on a single source introduces significant bias, particularly due to clinical recording settings. The data primarily captures specific states—such as resting state or pathological events—while complex cognitive tasks are notably absent. Consequently, the diversity of EEG representations in current models remains limited. A possible solution is to incorporate task-based EEG datasets shared via OpenNeuro (https://openneuro.org/), where most datasets conform to the Brain Imaging Data Structure (BIDS) format. This facilitates easier aggregation.

Another challenge in general-purpose EEG foundation models is the heterogeneity of EEG electrode configurations. Clinical EEG typically includes 19-21 channels based on the 10-20 system, while available electrodes in occipital or temporal areas are not always identical across clinical EEG data. Additionally, research setups often adopt between 32 and 128 channels. Furthermore, some high-density EEG systems (i.e. EGI) use geodesic sensor net configuration, where electrodes are evenly placed. The heterogeneity requires channel alignment adapted to input channel configuration of the downstream tasks. Our review found that half of the surveyed studies addressed this by designing models that accept variable channel inputs. Strategies include concatenating channels into a single sequence, padding empty channels, or utilizing specific spatial encoder modules to mitigate variability. Others remapped electrodes into a standardized space or used varying attention masks. No study specifically addressed the performance of channel alignment. Thus, there is no consensus on the optimal strategy for handling this heterogeneity while maintaining performance.

The current review identified the use of different model backbones in the field. Transformer architecture remains the predominant backbone in the field, though SSMs are emerging as an alternative implementation. FEMBA adopted MAMBA architecture, a special form of SSM. MAMBA is characterized by introducing the selection mechanism (Gu and Dao, 2024). It adaptively changes parameters for input and output projections, instead of using time-invariant parameters in conventional SSMs. This specific mechanism allows MAMBA to capture the state-dependent and non-stationary nature of neural signals including EEG. Code Brain and Knowledge-guided S4 adopted the S4 architecture (Gu et al., 2022). It utilizes FFT convolution to capture long-context inputs. However, the S4 system assumes that dynamics are constant over time. A direct comparison between these two architectures is awaited in future research.

A key advantage of SSMs including MAMBA and S4 over transformers is their computational complexity: SSMs scale linearly with sequence length ($O(L)$), whereas Transformers scale quadratically ($O(L^2)$) (Gu and Dao, 2024; Guo et al., 2025; Yang and Jia, 2025). This efficiency is particularly critical for processing long EEG epochs or enabling real-time applications, such as brain computer interface, neurofeedback, or closed-loop brain stimulation.

Furthermore, capturing the spatial dependencies of neural representations is essential. Among the 19 surveyed models, 14 explicitly implemented spatial encoding mechanisms by using attention, learnable embeddings, or 2D convolutions. This highlights the importance of spatial topology in EEG modeling.

In terms of learning objectives, the first BENDR study adopted contrastive learning (Kostas et al., 2021), but recent research has shifted towards masked auto-encoding (MAE). Interestingly, despite the success of autoregressive models in LLMs (Raiaan et al., 2024), they are not popular in EEG. It remains open if future work will adopt this direction. Regarding model size, the number of trainable parameters varied widely, from approximately 3 million to 540 million. Critical questions remain regarding the appropriate model size for capturing whole-brain representations and the extent to which "scaling laws" (Kaplan et al., 2020), especially across models, apply to EEG data. Addressing these questions requires unified benchmarks to rigorously test scaling behaviors (Kastrati et al., 2025).

To evaluate the performance, studies employed a diverse range of downstream tasks. Motor imagery and emotion recognition were common for basic applications, while seizure detection, sleep staging, and artifact detection were standard for clinical applications. Currently, results cannot be compared across studies due to differences in data preprocessing (e.g., filtering and resampling) and evaluation metrics. While intra-study comparisons are permissible, the field urgently requires a unified benchmark suite. Finally, regarding fine-tuning for downstream tasks, the majority of studies (approx. two-thirds) relied on full-parameter fine-tuning, while only a few utilized parameter-efficient methods (e.g., adapters or updating only the classification head). Furthermore, all but the ALFEE study (Xiong et al., 2025) performed single-task fine-tuning, suggesting that we have not yet achieved a truly robust, general-purpose "zero-shot" EEG foundation model.

In conclusion, we reviewed the state of EEG foundation models as of July 2025. Our survey revealed significant heterogeneity in pretraining datasets, model backbones, and SSL objectives. While the field shows promise, current models still require substantial post-training adaptation for downstream tasks. Future progress will depend on resolving open issues in data preprocessing and establishing standardized benchmarks to ensure reliable model comparison.

Table 1 Foundation models included in the review and training dataset characteristics

| | Year, Author | Model Name | Pretraining Datasets | Data size | # Channels | Filtering in pretraining | Sampling Rate | Input length | Input feature domain |
|---|---|---|---|---|---|---|---|---|---|
| 1 | 2021, Kostas | BENDR | TUEG | > 1.5 TB | 19 | LP 120 Hz | 256 Hz | 60 seconds | Raw EEG signals |
| 2 | 2023, Yang | BIOT | SHHS, PREST, Cardiology (ECG) + others | n.a. | Variable | n.a. | 200 Hz | Variable | EEG patch (T) |
| 3 | 2024, Cui | Neuro-GPT | TUH EEG corpus | 5656 hrs | 22 | BP 0.5-100 Hz | 250 Hz | 64 seconds | Raw EEG signals |
| 4 | 2024, Jiang | LaBraM | 20 datasets | > 2500 hrs | Variable | BP 0.1-75 Hz | 200 Hz | Variable | EEG patch (T) |
| 5 | 2024, Kommineni | Knowledge-guided S4 | TUEG | > 16000 hrs | 19 | BP 0.5-50 Hz | 250 Hz | 60 seconds | Raw EEG signals |
| 6 | 2024, Li | TSP | SEED series, EH, TUAB, HBN, TDBRAIN, PRED+CT | n.a. | Variable | LP 75 Hz | 200 Hz | 10 seconds | EEG patch (F) |
| 7 | 2024, Wang | EEGPT | PhysioMI, HGD, TSU, SEED, M3CV | n.a. | 58 | LP 38 Hz | 256 Hz | Variable | EEG patch (T) |
| 8 | 2024, Wu | EEG-ARNet | MOABB (120k samples) | n.a. | 128 | BP 5-95 Hz | n.a. | 512 points | Raw EEG signals |
| 9 | 2025, Bettinardi | BioSerenity-E1 | Proprietary + TUEG | 4005 hrs | 16 | BP 0.1-100 Hz | 128 Hz | 16 seconds | EEG patch (F) |
| 10 | 2025, Dimofte | CEReBrO | TUEG (w/o TUAB) | 20000 hrs | Variable (18/36) | n.a. | 256 Hz | Variable | EEG patch (T) |
| 11 | 2025, Feng | M4CEA | CHZU EEG Corpus | 1092 hrs | 16 | BP 0.5-70 Hz | 200 or 500Hz | Variable | EEG patch (T) |
| 12 | 2025, Li | Gram | 15 datasets | > 7000 hrs | Variable | n.a. | 200 Hz | Variable | EEG patch (T) |
| 13 | 2025, Liu | CRIA | TUEG | n.a. | Variable | BP 0.1-75 Hz | 200 Hz | Variable | EEG patch (T,F) |
| 14 | 2025, Ma | CodeBrain | TUEG | 9246 hrs | 19 | BP 0.5-100 Hz | 200 Hz | 10 seconds | EEG patch (T,F) |
| 15 | 2025, Ogg | HuBERT-based | TUEG | >24000 hrs | 8 | BP 0.1-50 Hz | 125 Hz | 60 seconds | Raw EEG signals |
| 16 | 2025, Tegon | FEMBA | TUEG | >21000 hrs | Variable | BP 0.1-100 Hz | n.a. | ~5-6 seconds | Raw EEG signals |
| 17 | 2025, Wang | CBraMod | TUEG | > 9000 hrs | 19 | BP 0.3-75 Hz | 200 Hz | 30 seconds | EEG patch (T,F) |
| 18 | 2025, Xiong | ALFEE | 15 datasets | 27062 hrs | Variable | BP 0.5-100 Hz | 256 Hz | Variable (2s -10s) | Raw EEG signals |
| 19 | 2025, Zhang | DMAE-EEG | 14 public datasets | n.a. | 10 super-nodes | n.a. | n.a. | Variable | EEG patch (T) |

Table 2 Model architecture and pretraining objectives

| | Model Name | Trainable Parameters | Model backbone | Attention type | Spatial Information Learning | Self-Supervised Task | Loss Function Type | Two-stage training |
|---|---|---|---|---|---|---|---|---|
| 1 | BENDR | n.a. (from Repo 157M) | 1D CNN + Transformer | Temporal | Channel-agnostic | CL | Contrastive Loss | |
| 2 | BIOT | 3.3M | Transformer | Joint spatiotemporal | Learnable channel embedding | CL | Cross-Entropy | |
| 3 | Neuro-GPT | n.a. | 2D CNN + Transformer | Temporal | Learnable 2D Convolution | Auto-regressive | MSE | |
| 4 | LaBraM | 5.8 - 369M | VQ Tokenizer + Transformer | Joint spatiotemporal | Learnable channel embedding | Masked Recon (cont) + Masked Token (dicrete) | MSE+Cross-Entropy | Yes |
| 5 | Knowledge-guided S4 | 13M | SSM (S4) | no | Channel-agnostic | Masked Recon (cont) + EEG power | Cosine + L1 | |
| 6 | TSP | n.a. | 2D CNN + Transformer | Spatial attention | Fixed channel embedding + spatial transformers | Masked Recon (cont) + CL | MSE + InfoNCE | |
| 7 | EEGPT | 4.7 - 25 M | Vision Transformer | Joint spatiotemporal | Learnable channel embedding | Spatio-temporal Representation + Masked Recon (cont) | MSE (Composite) | |
| 8 | EEG-ARNet | n.a. | 1D CNN + Transformer | Temporal | Channel-agnostic | Auto-regressive | MSE | |
| 9 | BioSerenity-E1 | 11.7M | VQ Tokenizer + Transformer | Joint spatiotemporal | Fixed channel embedding | Masked Recon (cont) + Masked Token (dicrete) | MSE+Cross-Entropy | Yes |
| 10 | CEReBrO | 3.58 - 85.15M | Transformer | Separated spatiotemporal | Learnable channel embedding | Masked Recon (cont) | MSE | |
| 11 | M4CEA | 5.8M - 6.3M | Vision Transformer | Temporal | Channel-agnostic | Masked Recon (cont) | MSE | |
| 12 | Gram | 6.0 - 251.28M | VQ Tokenizer + Transformer | Joint spatiotemporal | Fixed channel embedding | Masked Recon (cont) + Masked Token (dicrete) | Cosine + (Cross-Entropy + MSE) | Yes |
| 13 | CRIA | n.a. | Transformer | Separated spatiotemporal | Learnable channel embedding | CL | Cross-Entropy | |
| 14 | CodeBrain | 15.17M | VQ Tokenizer + Transformer + SSM (S4) | Temporal | Region-Based / Head Topological | Masked Recon (cont) + Masked Token (dicrete) | MSE+Cross-Entropy | Yes |
| 15 | HuBERT-based | 96.4M | 1D CNN + Transformer | Temporal | Channel-agnostic | k-means clustering + Masked Token (dicrete) | kluster sum of squares + Cross-Entropy | Yes |
| 16 | FEMBA | 7.8 - 389M | SSM (MAMBA) | no | Learnable 2D Convolution + leanable channel embeddings | Masked Recon (cont) | MSE | |
| 17 | CBraMod | 4.0M | Transformer | Separated spatiotemporal | Learnable 2D Convolution | Masked Recon (cont) | MSE | |
| 18 | ALFEE | 16.3 - 540M | 1D CNN + Channel Encoder + Temporal Transformer | Separated spatiotemporal | Learnable channel embedding | Auto-regressive+ Masked Recon (cont) + Task-specific classification | MSE + MSE + Cross-Entropy | |
| 19 | DMAE-EEG | n.a. | Vision Transformer | Joint spatiotemporal | Region-Based / Head Topological | Masked Recon (cont) | MSE | |

Table 3 Finetuning strategy for downstream tasks and code availability

| | Model Name | Finetuning strategy 1 | Outcomes | Finetuning strategy 2 | Downstream Tasks | Code availability |
|---|---|---|---|---|---|---|
| 1 | BENDR | Full, linear probing, | full ~= linear probing | Single-task finetuning | MI, ERN, P300, Sleep | Yes |
| 2 | BIOT | Full parameter tuning | | Single-task finetuning | Seizure, Abnormal/Event Detection, Arrhythmias (ECG), HAR | Yes |
| 3 | Neuro-GPT | Encoder only, Full, linear probing, | encoder > linear > full | Single-task finetuning | MI | Yes |
| 4 | LaBraM | Full, last 4, 8,12 blocks, linear probing | Full > 8block > linear | Single-task finetuning | Abnormal/Event Detection, ER, Gait Prediction | Yes |
| 5 | Knowledge-guided S4 | All S4 layer, last S4 layer, linear probe | Full > last S4 > linear | Single-task finetuning | Motor Movement, MI | |
| 6 | TSP | Full | | Single-task finetuning | ER, Event Detection | |
| 7 | EEGPT | Linear probing | | Single-task finetuning | MI, Sleep, ERN, P300, Abnormal/Event Detection | Yes |
| 8 | EEG-ARNet | Full, linear probing, | full > linear | Single-task finetuning | Visual Classification | |
| 9 | BioSerenity-E1 | Linear probing | | Single-task finetuning | Seizure, Abnormal Classification | |
| 10 | CEReBrO | Full parameter tuning | | Single-task finetuning | Anomaly/Seizure/ER/Gait | |
| 11 | M4CEA | Full, linear probing, | full > linear | Single-task finetuning | Artifact/Onset/Seizure Detection, Syndrome Classification, etc. | Yes |
| 12 | Gram | Full | | Single-task finetuning | ER, Event Classification, Sleep, Restoration | Yes |
| 13 | CRIA | Full parameter tuning | | Single-task finetuning | Abnormal/Event Detection, Seizure | |
| 14 | CodeBrain | Linear probing | | Single-task finetuning | ER, Sleep, MI, Seizure, etc. | |
| 15 | HuBERT-based | Full, linear probing, | full > linear | Single-task finetuning | P300, MI, RSVP, Participant ID | |
| 16 | FEMBA | Full parameter tuning | | Single-task finetuning | Abnormal/Artifact/Slowing Detection | |
| 17 | CBraMod | Full, linear probing | Full > linear | Single-task finetuning | ER, MI, Sleep, Seizure Detection, Abnormal/Event Detection | Yes |
| 18 | ALFEE | Full parameter tuning | | Multi-task finetuning | Abnormal/Event Detection, Seizure, ER, Sleep, Workload, MI | |
| 19 | DMAE-EEG | Linear probing (except enhancement task) | | Single-task finetuning | Signal Quality Enhancement, Motion Intention Recognition | |

References

Abibullaev, B., Keutayeva, A., and Zollanvari, A. (2023). Deep Learning in EEG-Based BCIs: A Comprehensive Review of Transformer Models, Advantages, Challenges, and Applications. *IEEE Access* 11, 127271–127301. doi: 10.1109/ACCESS.2023.3329678

Abiri, R., Borhani, S., Sellers, E. W., Jiang, Y., and Zhao, X. (2019). A comprehensive review of EEG-based brain–computer interface paradigms. *J. Neural Eng.* 16, 011001. doi: 10.1088/1741-2552/aaf12e

Aggarwal, S., and Chugh, N. (2022). Review of Machine Learning Techniques for EEG Based Brain Computer Interface. *Arch Computat Methods Eng* 29, 3001–3020. doi: 10.1007/s11831-021-09684-6

Albelwi, S. (2022). Survey on Self-Supervised Learning: Auxiliary Pretext Tasks and Contrastive Learning Methods in Imaging. *Entropy* 24, 551. doi: 10.3390/e24040551

Berger, H. (1929). Über das elektrenkephalogramm des menschen. *European Archives of Psychiatry and Clinical Neuroscience* 87, 527–570.

Bettinardi, R. G., Rahmouni, M., and Gimenez, U. (2025). BioSerenity-E1: a self-supervised EEG model for medical applications. doi: 10.48550/arXiv.2503.10362

Bommasani, R., Hudson, D. A., Adeli, E., Altman, R., Arora, S., Arx, S. von, et al. (2022). On the Opportunities and Risks of Foundation Models. doi: 10.48550/arXiv.2108.07258

Brigo, F., and Mecarelli, O. (2025). *EEG: The First 100 Years: Past, Present and Future of Electroencephalography*. Springer Nature.

Chen, X., He, J., Wu, X., Yan, W., and Wei, W. (2020). Sleep staging by bidirectional long short-term memory convolution neural network. *Future Generation Computer Systems* 109, 188–196. doi: 10.1016/j.future.2020.03.019

Craik, A., He, Y., and Contreras-Vidal, J. L. (2019). Deep learning for electroencephalogram (EEG) classification tasks: a review. *Journal of neural engineering* 16, 031001.

Cui, W., Jeong, W., Thölke, P., Medani, T., Jerbi, K., Joshi, A. A., et al. (2024). Neuro-GPT: Towards A Foundation Model For EEG., in *2024 IEEE International Symposium on Biomedical Imaging (ISBI)*, 1–5. doi: 10.1109/ISBI56570.2024.10635453

Dimofte, A., Bucagu, G. A., Ingolfsson, T. M., Wang, X., Cossettini, A., Benini, L., et al. (2025). CEReBrO: Compact Encoder for Representations of Brain Oscillations Using Efficient Alternating Attention. doi: 10.48550/arXiv.2501.10885

Ding, X., and He, Q. (2017). Energy-Fluctuated Multiscale Feature Learning With Deep ConvNet for Intelligent Spindle Bearing Fault Diagnosis. *IEEE Transactions on Instrumentation and Measurement* 66, 1926–1935. doi: 10.1109/TIM.2017.2674738

Efe, E., and Ozsen, S. (2023). CoSleepNet: Automated sleep staging using a hybrid CNN-LSTM network on imbalanced EEG-EOG datasets. *Biomedical Signal Processing and Control* 80, 104299. doi: 10.1016/j.bspc.2022.104299

Erat, K., Şahin, E. B., Doğan, F., Merdanoğlu, N., Akcakaya, A., and Durdu, P. O. (2024). Emotion recognition with EEG-based brain-computer interfaces: a systematic literature review. *Multimed Tools Appl* 83, 79647–79694. doi: 10.1007/s11042-024-18259-z

Feng, Y., Hu, D., Jiang, T., Gao, F., and Cao, J. (2025). M4CEA: A Knowledge-guided Foundation Model for Childhood Epilepsy Analysis. *IEEE J Biomed Health Inform* PP. doi: 10.1109/JBHI.2025.3590463

Gu, A., and Dao, T. (2024). Mamba: Linear-Time Sequence Modeling with Selective State Spaces. doi: 10.48550/arXiv.2312.00752

Gu, A., Goel, K., and Ré, C. (2022). Efficiently Modeling Long Sequences with Structured State Spaces. doi: 10.48550/arXiv.2111.00396

Guo, M., Han, X., Liu, H., Zhu, J., Zhang, J., Bai, Y., et al. (2025). MI-Mamba: A hybrid motor imagery electroencephalograph classification model with Mamba's global scanning. *Annals of the New York Academy of Sciences* 1544, 242–253. doi: 10.1111/nyas.15288

Hochreiter, S., and Schmidhuber, J. (1997). Long Short-Term Memory. *Neural Computation* 9, 1735–1780. doi: 10.1162/neco.1997.9.8.1735

Jaipriya, D., and Sriharipriya, K. C. (2024). Brain Computer Interface-Based Signal Processing Techniques for Feature Extraction and Classification of Motor Imagery Using EEG: A Literature Review. *Biomedical Materials & Devices* 2, 601–613. doi: 10.1007/s44174-023-00082-z

Jiang, W.-B., Zhao, L.-M., and Lu, B.-L. (2024). Large Brain Model for Learning Generic Representations with Tremendous EEG Data in BCI. doi: 10.48550/arXiv.2405.18765

Kaplan, J., McCandlish, S., Henighan, T., Brown, T. B., Chess, B., Child, R., et al. (2020). Scaling Laws for Neural Language Models. doi: 10.48550/arXiv.2001.08361

Kashefi Amiri, H., Zarei, M., and Daliri, M. R. (2025). Epileptic seizure detection from electroencephalogram signals based on 1D CNN-LSTM deep learning model using discrete wavelet transform. *Sci Rep* 15, 32820. doi: 10.1038/s41598-025-18479-9

Kastrati, A., Bürki, J., Lauer, J., Xuan, C., Iaquinto, R., and Wattenhofer, R. (2025). EEG-Bench: A Benchmark for EEG Foundation Models in Clinical Applications. *arXiv.org*. Available at: https://arxiv.org/abs/2512.08959v1 (Accessed December 16, 2025).

Kommineni, A., Avramidis, K., Leahy, R., and Narayanan, S. (2024). Knowledge-guided EEG Representation Learning., in *2024 46th Annual International Conference of the IEEE Engineering in Medicine and Biology Society (EMBC)*, 1–6. doi: 10.1109/EMBC53108.2024.10782310

Kostas, D., Aroca-Ouellette, S., and Rudzicz, F. (2021). BENDR: Using transformers and a contrastive self-supervised learning task to learn from massive amounts of EEG data. *Frontiers in Human Neuroscience* 15, 653659.

Krizhevsky, A., Sutskever, I., and Hinton, G. E. (2012). Imagenet classification with deep convolutional neural networks. *Advances in neural information processing systems* 25. Available at: https://proceedings.neurips.cc/paper/2012/hash/c399862d3b9d6b76c8436e924a68c45b-Abstract.html (Accessed December 14, 2025).

Lawhern, V. J., Solon, A. J., Waytowich, N. R., Gordon, S. M., Hung, C. P., and Lance, B. J. (2018). EEGNet: a compact convolutional neural network for EEG-based brain–computer interfaces. *Journal of neural engineering* 15, 056013.

Li, Z., Zhao, L.-M., Zheng, W.-L., and Lu, B.-L. (2024). Temporal-Spatial Prediction: Pre-Training on Diverse Datasets for EEG Classification., in *ICASSP 2024 - 2024 IEEE International Conference on Acoustics, Speech and Signal Processing (ICASSP)*, 1806–1810. doi: 10.1109/ICASSP48485.2024.10447845

Li, Z., Zheng, W.-L., and Lu, B.-L. (2025). Gram: A Large-Scale General EEG Model for Raw Data Classification and Restoration Tasks., in *ICASSP 2025 - 2025 IEEE International Conference on Acoustics, Speech and Signal Processing (ICASSP)*, 1–5. doi: 10.1109/ICASSP49660.2025.10890831

Liu, P., Chen, C. L. P., He, Y., and Zhang, T. (2025). CRIA: A Cross-View Interaction and Instance-Adapted Pre-training Framework for Generalizable EEG Representations. doi: 10.48550/arXiv.2506.16056

Liu, X., Zhang, F., Hou, Z., Mian, L., Wang, Z., Zhang, J., et al. (2023). Self-Supervised Learning: Generative or Contrastive. *IEEE Transactions on Knowledge and Data Engineering* 35, 857–876. doi: 10.1109/TKDE.2021.3090866

Lotte, F., Bougrain, L., Cichocki, A., Clerc, M., Congedo, M., Rakotomamonjy, A., et al. (2018). A review of classification algorithms for EEG-based brain–computer interfaces: a 10 year update. *J. Neural Eng.* 15, 031005. doi: 10.1088/1741-2552/aab2f2

Ma, J., Wu, F., Lin, Q., Xing, Y., Liu, C., Jia, Z., et al. (2025). CodeBrain: Towards Decoupled Interpretability and Multi-Scale Architecture for EEG Foundation Model. doi: 10.48550/arXiv.2506.09110

Obeid, I., and Picone, J. (2016). The Temple University Hospital EEG Data Corpus. *Front. Neurosci.* 10. doi: 10.3389/fnins.2016.00196

Ogg, M., Hingorani, R., Luna, D., Milsap, G. W., Coon, W. G., and Scholl, C. A. (2025). EEG Foundation Models for BCI Learn Diverse Features of Electrophysiology. doi: 10.48550/arXiv.2506.01867

Rafiei, M. H., Gauthier, L. V., Adeli, H., and Takabi, D. (2024). Self-Supervised Learning for Electroencephalography. *IEEE Transactions on Neural Networks and Learning Systems* 35, 1457–1471. doi: 10.1109/TNNLS.2022.3190448

Raiaan, M. A. K., Mukta, M. S. H., Fatema, K., Fahad, N. M., Sakib, S., Mim, M. M. J., et al. (2024). A review on large language models: Architectures, applications, taxonomies, open issues and challenges. *IEEE access* 12, 26839–26874.

Rani, V., Nabi, S. T., Kumar, M., Mittal, A., and Kumar, K. (2023). Self-supervised Learning: A Succinct Review. *Arch Computat Methods Eng* 30, 2761–2775. doi: 10.1007/s11831-023-09884-2

Rashid, M., Sulaiman, N., P. P. Abdul Majeed, A., Musa, R. M., Ab. Nasir, A. F., Bari, B. S., et al. (2020). Current Status, Challenges, and Possible Solutions of EEG-Based Brain-Computer Interface: A Comprehensive Review. *Front. Neurorobot.* 14. doi: 10.3389/fnbot.2020.00025

Roy, Y., Banville, H., Albuquerque, I., Gramfort, A., Falk, T. H., and Faubert, J. (2019). Deep learning-based electroencephalography analysis: a systematic review. *Journal of neural engineering* 16, 051001.

Saha, S., and Baumert, M. (2020). Intra- and Inter-subject Variability in EEG-Based Sensorimotor Brain Computer Interface: A Review. *Front. Comput. Neurosci.* 13. doi: 10.3389/fncom.2019.00087

Salisbury, D. F., Fisher, D., and Di Lorenzo, G. (2024). Editorial: 100th year anniversary of the discovery of electroencephalography. *Clin EEG Neurosci* 55, 3–3. doi: 10.1177/15500594231217519

Tegon, A., Ingolfsson, T. M., Wang, X., Benini, L., and Li, Y. (2025). FEMBA: Efficient and Scalable EEG Analysis with a Bidirectional Mamba Foundation Model. doi: 10.48550/arXiv.2502.06438

Vafaei, E., and Hosseini, M. (2025). Transformers in EEG Analysis: A Review of Architectures and Applications in Motor Imagery, Seizure, and Emotion Classification. *Sensors* 25, 1293. doi: 10.3390/s25051293

Vaswani, A., Shazeer, N., Parmar, N., Uszkoreit, J., Jones, L., Gomez, A. N., et al. (2017). Attention is all you need. *Advances in neural information processing systems* 30.

Wang, G., Liu, W., He, Y., Xu, C., Ma, L., and Li, H. (2024). EEGPT: Pretrained transformer for universal and reliable representation of eeg signals. *Advances in Neural Information Processing Systems* 37, 39249–39280.

Wang, J., Zhao, S., Luo, Z., Zhou, Y., Jiang, H., Li, S., et al. (2025). CBraMod: A Criss-Cross Brain Foundation Model for EEG Decoding. doi: 10.48550/arXiv.2412.07236

Wei, C.-S., Lin, Y.-P., Wang, Y.-T., Lin, C.-T., and Jung, T.-P. (2018). A subject-transfer framework for obviating inter- and intra-subject variability in EEG-based drowsiness detection. *NeuroImage* 174, 407–419. doi: 10.1016/j.neuroimage.2018.03.032

Weng, W., Gu, Y., Guo, S., Ma, Y., Yang, Z., Liu, Y., et al. (2025). Self-supervised Learning for Electroencephalogram: A Systematic Survey. *ACM Comput. Surv.* 57, 317:1-317:38. doi: 10.1145/3736574

Wu, A., Zhang, Y., Yu, Y., and Zeng, L.-L. (2024). EEG-ARNet: An Autoregressive Pre-training Model for Extracting EEG Features., in *2024 5th International Conference on Computers and Artificial Intelligence Technology (CAIT)*, 255–259. doi: 10.1109/CAIT64506.2024.10962980

Xiong, W., Lin, J., Li, J., Li, J., and Jiang, C. (2025). ALFEE: Adaptive Large Foundation Model for EEG Representation. doi: 10.48550/arXiv.2505.06291

Xu, G., Ren, T., Chen, Y., and Che, W. (2020). A One-Dimensional CNN-LSTM Model for Epileptic Seizure Recognition Using EEG Signal Analysis. *Front. Neurosci.* 14. doi: 10.3389/fnins.2020.578126

Yang, C., Westover, M. B., and Sun, J. (2023). BIOT: Cross-data Biosignal Learning in the Wild. doi: 10.48550/arXiv.2305.10351

Yang, X., and Jia, Z. (2025). Spatial-Temporal Mamba Network for EEG-Based Motor Imagery Classification., 418–432. doi: 10.1007/978-981-96-0821-8_28

Zhang, Y., Yu, Y., Li, H., Wu, A., Chen, X., Liu, J., et al. (2025). DMAE-EEG: A Pretraining Framework for EEG Spatiotemporal Representation Learning. *IEEE Transactions on Neural Networks and Learning Systems* 36, 17664–17678. doi: 10.1109/TNNLS.2025.3581991

Zhu, L., Guan, Q., Liu, Y., Xu, J., and Ma, S. (2025). Research on Sleep Stage Classification of Electroencephalogram Signals Based on CNN and LSTM. *IEEE Access* 13, 75351–75362. doi: 10.1109/ACCESS.2025.3559350